\documentclass[12pt,showpacs,preprintnumbers,amsmath,amssymb,nofootinbib,superscriptaddress]{article}
\usepackage{graphicx}
\usepackage{dcolumn}
\usepackage{bm}
\usepackage{graphics}
\usepackage{amssymb}
\usepackage{amscd}
\usepackage{afterpage}
\usepackage{float,times}
\usepackage{subfigure}
\usepackage{rotating}
\usepackage{multirow}
\usepackage{fancyheadings}
\usepackage{theorem}
\usepackage{moreverb}
\usepackage{euscript}
\usepackage{psfrag}
\begin{document}
\begin{titlepage}

%\begin{center}
%{\hbox to\hsize{
%\hfill \bf hep-ph/??? }}
{\hbox to\hsize{\hfill November 2012 }}

\bigskip \vspace{3\baselineskip}

\begin{center}
{\bf \large 
Time-symmetric quantization in spacetimes with event horizons}

\bigskip

\bigskip

{\bf Archil Kobakhidze$^{\rm a}$ and Nicholas L. Rodd$^{\rm b}$  \\}

\smallskip

{ \small \it
ARC Centre of Excellence for Particle Physics at the Terascale, \\
$^{\rm a}$ School of Physics, The University of Sydney, NSW 2006, Australia \\ 
$^{\rm b}$ School of Physics, The University of Melbourne, VIC 3010, Australia\\
E-mail: archilk@physics.usyd.edu.au; nrodd@student.unimelb.edu.au
\\}

\bigskip
 
\bigskip

\bigskip

{\large \bf Abstract}

\end{center}
%\noindent 
{\small
The standard quantization formalism in spacetimes with event horizons implies a non-unitary evolution of quantum states, as initial pure states may evolve into thermal states.  This phenomenon is behind the famous black hole information loss paradox which provoked long-standing debates on the compatibility of quantum mechanics and gravity.  In this paper we demonstrate that within an alternative time-symmetric quantization formalism thermal radiation is absent and states evolve unitarily in spacetimes with event horizons. We also discuss the theoretical consistency of the proposed formalism. We explicitly demonstrate that the theory preserves the microcausality condition and suggest a ``reinterpretation postulate" to resolve other apparent pathologies associated with negative energy states. Accordingly as there is a consistent alternative, we argue that choosing to use time-asymmetric quantization is a necessary condition for the black hole information loss paradox.}

\vspace{1cm}
\end{titlepage}

\section{Black hole information paradox} 

The prediction of thermal radiation from classical black holes \cite{Hawking:1974rv} has ignited long-standing debates on the compatibility of quantum mechanics and gravity. The standard quantization of fields in the background spacetime of a classical black hole implies that a pure quantum mechanical state evolves into a mixed state of Hawking radiation, thereby destroying all the information about the initial quantum state. This violates the basic postulate of quantum mechanics according to which quantum mechanical probabilities are conserved during the evolution of a system, which means that information related to the system is necessarily conserved. 

The black hole information paradox may well be an artefact of the approximation where gravity is kept classical. Indeed, there are some suggestions that the paradox is resolved within the more fundamental string theory, for example using the  AdS/CFT correspondence \cite{Aharony:1999ti} or more specific ``fuzzball'' descriptions of black hole microstates \cite{Mathur:2005zp}. Another interesting scenario was recently discussed in \cite{Dvali:2011aa}.

In addition to this, however, it is important to understand exactly how the effect emerges at the level where gravity is kept classical. It is already well known that thermal radiation appears as a quantum mechanical effect even in flat spacetimes without gravity (Unruh radiation \cite{Fulling:1972md}), suggesting that Hawking radiation is a purely kinematic effect (see also the discussion in \cite{Visser:2001kq}). The aim of this paper is to demonstrate that the black hole information paradox results from not only the peculiar structure of spacetimes with event horizons, but further due to the time-asymmetric nature of the standard quantization of fields.

Within the standard quantization a relativistic field operator in a given inertial reference frame is broken up into positive and negative frequency parts, which are associated with annihilation and creation operators for positive energy states only. The restriction to only positive energy states has the merit that we always deal with positive norm states and causal propagation as given by the Feynman propagator. Nevertheless in accelerated (non-inertial) reference frames in flat Minkowski spacetime or in a black hole spacetime where no global inertial reference frames are available, this restriction to positive energy states leads inevitably to the existence of thermal states with respect to the Killing vector of time translations. If instead a time-symmetric quantization is used, where annihilation and creation operators are associated with both positive and negative states, the negative energy states lead to a cancelling out of the thermal effects. Thus in this formalism there is no black hole information paradox as Hawking radiation has been removed, revealing the time-asymmetric quantization as a fundamental condition for its existence.

The remainder of the paper is structured as follows. Section 2 outlines the time-symmetric quantization and canvasses its key features. It is shown that with an appropriate choice of metric all states have a positive norm and also that the theory does not violate causality. In Section 3 we demonstrate that such a quantization leads to the absence of thermal radiation in spacetimes with event horizons within the technically simpler Rindler spacetime, which captures all the essential ingredients of time-symmetric quantization in black hole spacetimes, and thus the formalism can be used equally well to demonstrate the absence of Hawking radiation. In Section 4 we briefly describe a ``reinterpretation principle", whereby the negative energy states can be reinterpreted as positive energy states in order to resolve pathologies with the formalism. Finally in Section 5 we discuss our results.

\section{Time-symmetric quantization} 

In this section we show how to introduce the time-symmetric quantization and consider its basic features. To begin with consider the quantization of a free massless scalar field in an inertial reference frame. We assume the standard quantization, so the field is expanded in the complete basis of positive and negative frequency plane-wave modes, 
\begin{equation}
\hat \phi(t,x)=\int dk \left(f_k\hat A_k+f_k^{*}\hat A_k^{\dagger}\right)~,
\label{1}
\end{equation}
where $f_k=\frac{1}{2\pi\sqrt{2|k|}}{\rm e}^{-i(|k|t-kx)}$ and rather than the traditional $\hat A_k = \hat a_k$, in order to ensure a time-symmetric quantization we take
\begin{equation}
\hat A_k=\frac{1}{\sqrt{2}}\left(\hat a_k+\hat \alpha_{-k}^{\dagger}\right)~,~\left[\hat A_k^{\dagger}\equiv(\hat A_k)^{\dagger}\right]~.
\label{2}
\end{equation}
This now ensures that $\hat \phi(t,x)$ has a symmetry $t \rightarrow -t$, an explicit time-symmetry as is present in the Klein-Gordon equation, rather than an anti-unitary one. The annihilation $\hat a_k$, $\hat \alpha_k$ and creation $\hat a_k^{\dagger}$, $\hat \alpha_k^{\dagger}$ operators act on an extended Hilbert space ${\cal H}^{(+)}\bigoplus{\cal H}^{(-)}$ (known as the Hilbert-Krein space)\footnote{Extension to Hilbert-Krein space has been proposed also in \cite{Gazeau:1999mi} to maintain a de Sitter covariant quantization.} and satisfy the following commutation relations:
 \begin{eqnarray}
 [\hat a_k, \hat a^{\dagger}_{k'}]=2\pi\delta(k-k') \\
\label{3}  
  [\hat \alpha_k,\hat \alpha^{\dagger}_{k'}]=-2\pi\delta(k-k')~,
   \label{4}
 \end{eqnarray}
 where the remaining commutators are trivial. Notice that $[\hat A_k, \hat A^{\dagger}_{k'}]=2\pi\delta(k-k')$. The vacuum state in the inertial reference frame, $\vert 0_M\rangle$, is defined as:
\begin{equation}
\hat a_k\vert 0_M\rangle=\hat \alpha_k\vert 0_M\rangle=0~,~~\forall k \in \mathbb{R}~.
\label{5}
\end{equation}
The inner product on the Hilbert-Krein space ${\cal H}^{(+)}\bigoplus{\cal H}^{(-)}$,
\begin{equation}
(\psi, \varphi)\equiv \langle \psi \vert \hat \eta \vert \varphi \rangle~( \geq 0)~,
\label{6}
\end{equation}
is defined with the aid of the metric-operator $\hat \eta$, which can be represented as:
\begin{equation}
\hat \eta = \left(-1\right)^{-\int \frac{dk}{2\pi} \hat \alpha_k^{\dagger}\hat \alpha_k}~.
\label{7}
\end{equation}
Then we can ensure a positive norm for an arbitrary multiparticle Fock state with $n_i$ $a$-particles of momenta $k_i$ and $m_i$ $\alpha$-particles of momenta $l_i$ - specifically: 
\begin{equation}
\vert n_1 k_1...n_i k_i...m_1 l_1...m_j l_j...\rangle=\prod_{\{ \lambda \}}\prod_{\{ \kappa \}}\frac{(\hat a^{\dagger}_{k_\lambda})^{n_\lambda} (\hat \alpha^{\dagger}_{l_\kappa})^{n_\kappa}}{\sqrt{n_\lambda !} \sqrt{m_\kappa !}} \vert 0_M\rangle ~,
\label{8}
\end{equation}
where $\{ \lambda \}$ and $\{ \kappa \}$ are the occupied states for the $a$ and $\alpha$-particles respectively. Thus a consistent probabilistic quantum mechanical description can be maintained in the theory. Despite this the negative energy states are still present and such states typically pose serious problems. We will discuss a possible resolution of such problems in Section 4.

The particle number operator is usually defined through the conserved charge (the Klein-Gordon product): 
\begin{equation}
\langle f, h \rangle=\int dx j^{0}= i\int dx \left(f^{*}\partial_t h-(\partial_t f^{*})h \right)~,
\label{9a}
\end{equation}
where $f$ and $h$ are solutions of the Klein-Gordon equation. Physically, the particle number operator is required to have positive eigenvalues, and therefore positive $\hat \phi^{+}$ and negative $\hat \phi^{-}$ frequency contributions to the particle number operator enter with opposite signs: 
\begin{equation}
\hat N = \frac{1}{2} : \left(\langle \hat \phi^{+} , \hat \phi^{+} \rangle - \langle \hat \phi^{-} , \hat \phi^{-} \rangle \right):~,
\label{10}
\end{equation} 
where $:~...~:$ is for normal ordering. For the traditional quantization $\hat \phi^{+}=\int dk f_k \hat a_k$, $\hat \phi^{-}=\int dk f_k^{*}\hat a_k^{\dagger}$ and we obtain:
\begin{equation}
\hat N_M = \frac{1}{2}:\int \frac{dk}{2\pi} \left[ \hat a^{\dagger}_k \hat a_k + \hat a_k \hat a^{\dagger}_k \right]:~= \int \frac{dk}{2\pi} \left[ \hat a^{\dagger}_k \hat a_k\right]~,
\label{11}
\end{equation} 
which is the usual result. Note in particular that normal ordering was necessary in order to prevent a divergent contribution $\sim \delta (0)$ appearing in the result. Proceeding instead with the time-symmetric field $\hat \phi^{+}=\int dk f_k \hat A_k$, $\hat \phi^{-}=\int dk f_k^{*}\hat A_k^{\dagger}$, we find:
\begin{equation}
\hat N_M =~:\int \frac{dk}{4\pi} \left(\hat A^{\dagger}_k\hat A_k+\hat A_k\hat A^{\dagger}_k\right):~=~:\int \frac{dk}{4\pi}\left(\hat a^{\dagger}_k\hat a_k+\hat \alpha^{\dagger}_k\hat \alpha_k +\hat a_k\hat \alpha_{-k}+ \hat a^{\dagger}_k\hat \alpha^{\dagger}_{-k}\right):~,
\label{12}
\end{equation}
where the absence of divergent contributions in the final result is due to a cancellation brought about by the opposite sign in the $\hat \alpha$ commutator (\ref{4}). Accordingly in this theory there is no need for normal ordering. Nevertheless this number operator has some unusual properties: (i) $\hat N_M$ is not positive definite, that is, $\hat N_M$ has negative expectation values evaluated for states containing $\alpha$-particles, for example, $\langle l_1 \vert \hat N_M\hat \eta \vert l_2 \rangle=
\langle 0_M\vert \hat \alpha_{l_1} \hat N_M\hat \eta \hat \alpha^{\dagger}_{l_2}\vert 0_M\rangle= - \pi \delta \left(l_1-l_2 \right) \langle 0_M\vert 0_M\rangle=-\pi \delta \left(l_1-l_2 \right)$; and (ii) the number operator is not diagonal in the oscillator basis introduced above, for instance $\langle 0_M\vert \hat N_M \hat \eta \vert k, l\rangle=-\langle 0_M\vert \hat N_M \hat a^{\dagger}_k \hat \alpha^{\dagger}_l \vert 0_M\rangle= \pi \delta(k+l)\neq 0$. The Hamiltonian of the system is expressed through the same combination of operators, 
\begin{equation}
\hat H=\int \frac{dk}{2\pi}\frac{|k|}{2} \left(\hat A^{\dagger}_k\hat A_k+\hat A_k\hat A^{\dagger}_k\right) 
\label{13}
\end{equation} 
and thus shares the same unusual properties as $\hat N_M$. In particular, the states (\ref{8}) are not eigenstates of the Hamiltonian (\ref{13}). Again, note there is no need for normal ordering, $\langle 0 \vert H \eta \vert 0 \rangle = 0$ automatically. In the free field case the vacuum energy is the only divergent quantity, therefore free time-symmetric field theory is a finite theory (in any dimensions).    

It is easy to confirm by direct calculation that the propagator for this theory is not the causal Feynman propagator $\Delta_F$, but rather a sum of half causal and half anti-causal [$\bar \Delta_F\equiv \Delta_F^{*}$] Feynman propagators:
%\begin{equation}
$\Delta_{WF}(\omega, k) = \frac{1}{2} \Delta_F(\omega, k) + \frac{1}{2} \bar \Delta_F(\omega, k)~.$
%\label{15}
%\end{equation}
Accordingly the on-shell part is absent, i.e.,
\begin{equation}
\Delta_{WF}(\omega, k)={\cal P}\frac{1}{\omega^2-k^2}~,
\label{14}
\end{equation}
where ${\cal P}$ denotes Cauchy principle value. The on-shell part is cancelled out between positive and negative energy contributions. Note that a similar propagator appears also in the action-at-a-distance formulation of classical electrodynamics by Wheeler and Feynman \cite{Wheeler:1945ps}. It would be interesting to see if there is a relation between the time-symmetric quantization formalism and a theory of non-local fields. Despite the fact that (\ref{14}) is not a causal Feynman propagator, we do not expect a violation of causality since the microcausality condition holds in our theory. Indeed, since the operators $\hat A_k$ and $\hat A^{\dagger}_k$ satisfy the standard commutation relation, one can obtain the standard result for the commutator of field operators:      
\begin{equation}
[\hat \phi(t,x), \hat \phi(t',x')]=-i\int\frac{dk}{2\pi |k|}\sin\left(|k|(t-t')-k(x-x')\right)~.
\label{15}
\end{equation}  
This vanishes for spacelike separated events $|t-t'|<|x-x'|$.

\section{Absence of Unruh radiation in an accelerated frame}

In this section we demonstrate the absence of thermal radiation for an observer in Rindler spacetime - the simplest spacetime with a horizon. The Rindler coordinate system describes a uniformly accelerating observer in (1+1)-dimensional flat Minkowski spacetime. The conformal Rindler coordinates $(\tau, \chi)$ can be written in terms of Cartesian coordinates $(t,x)$, which parametrize flat Minkowski spacetime, as follows:
\begin{equation}
\tau=\frac{1}{2a}\ln\left(\frac{x+t}{x-t} \right)~,~~\chi=\frac{1}{2a}\ln\left(a^2[x^2-t^2]\right)~,
\label{16}
\end{equation} 
where $a>0$ is the modulus of a constant acceleration. Accordingly the Rindler coordinates cover only the $x>|t|$ quadrant of the full Minkowski spacetime. The field quantization in this frame is formally similar to the one described in Section 2. The mode expansion reads:
\begin{equation}
\hat \phi(\tau,\chi)=\int dp \left(F_p\hat B_p+F_p^{*}\hat B_p^{\dagger}\right)~,
\label{17}
\end{equation}
where $F_p=\frac{1}{2\pi\sqrt{2|p|}}{\rm e}^{-i(|p|\tau-p\chi)}$ and 
\begin{equation}
\hat B_p=\frac{1}{\sqrt{2}}\left(\hat b_p+\hat \beta_{-p}^{\dagger}\right)~,~\left[\hat B_p^{\dagger}\equiv(\hat B_p)^{\dagger}\right]~.
\label{18} 
\end{equation}
However, an important difference is that $\lbrace F_{p}\rbrace$ form complete set only in the region $\vert x\vert>t$. Again the operator $\hat b_p$ ($\hat b_p^{\dagger}$) is interpreted as the annihilation (creation) operator for positive energy particles with momentum $p$ while the operator $\hat \beta_p$ ($\hat \beta_p^{\dagger}$) is the annihilation (creation) operators for negative energy particles with the same momentum. Non-trivial commutators read:
\begin{eqnarray}
\left[\hat b_p, \hat b_{p'}^{\dagger}\right]=2\pi\delta(p-p') \\
\left[\hat \beta_p, \hat \beta_{p'}^{\dagger}\right]=-2\pi\delta(p-p')~ ,
\label{19} 
\end{eqnarray}
and $[\hat B_p,\hat B^{\dagger}_{p'}]=2\pi \delta(p-p')$. The vacuum state $\vert 0_R \rangle$ in the accelerated reference frame is defined as: 
\begin{equation}
\hat b_p \vert 0_R \rangle = \hat \beta_p \vert 0_R \rangle =0~, ~~\forall p \in \mathbb{R}~.  
\label{20}
\end{equation}
We again see that Hilbert-Krein space must be equipped with an indefinite metric: $\hat \xi=(-1)^{-\int \frac{dp}{2\pi} \hat \beta^{\dagger}_p\hat \beta_p}$. The particle number operator has a form analogous to (\ref{12}):
\begin{equation}
\hat N_R =\int \frac{dp}{4\pi} \left(\hat B^{\dagger}_p\hat B_p+\hat B_p\hat B^{\dagger}_p\right)~.
\label{21}
\end{equation} 

Let us demonstrate now the absence of thermal radiation in the accelerated reference frame within the time-symmetric quantization. Because the operators $\hat A_k$ and $\hat B_p$ satisfy the standard commutation relations, we can simply borrow the usual Bogoliubov transformations which relates operators in inertial and accelerated reference frames \cite{Fulling:1972md}. Actually, we only need a generic form of these transformations:

\begin{equation}
\hat B_p=\int dk \left(\gamma_{pk}\hat A_k+\delta_{pk}\hat A_{k}^{\dagger}\right)~.
\label{22}
\end{equation}
The expectation value of the particle number operator (\ref{21}) in the Minkowski vacuum then is:
\begin{equation}
\langle 0_M\vert \hat N_R\vert 0_M\rangle = \int \frac{dpdldk}{4\pi} \left(\gamma^{*}_{pl}\gamma_{pk}+\delta^{*}_{pl}\delta_{pk}\right) 
\langle 0_M\vert \hat \alpha_{-l} \hat \alpha^{\dagger}_{-k} + \hat a_l \hat a^{\dagger}_k \vert 0_M\rangle = 0~,
\label{23}
\end{equation}

where in the second equality we have kept only terms that do not annihilate the vacuum and in the third we have made use of the commutation relations in (3) and (\ref{4}). The negative and positive frequency contributions cancel out in (\ref{23}) and thus an accelerated observer sees a non-thermal,  particle-free Minkowski vacuum, which is identical to what is seen by the inertial observer. In this sense the vacuum state is invariant under Bogoliubov transformations, as has been pointed out earlier in \cite{Garidi:2005}. This is to be contrasted with the result obtained via the standard quantization, where an accelerated Rindler observer sees the Minkowski vacuum as a thermal spectrum of particles. A similar analysis can be performed in Schwarzschild spacetime to show the absence of Hawking radiation within the framework of time-symmetric quantization.  

\section{Negative energy states and the reinterpretation postulate}

The negative energy states introduced in the time-symmetric quantization lead to an apparent stability problem since the energy is not bounded from below. For example, the vacuum state is apparently unstable since it may spontaneously decay into negative and positive energy states. Here, following earlier work on quantized tachyonic fields \cite{Dhar:1969gb}, we suggest a reinterpretation principle that resolves the issue with negative energy states. 

A transition amplitude of an initial state $\vert\Psi(t_{\rm in})\rangle$ into a final state  $\vert\Psi(t_{\rm f})\rangle$ is defined by the evolution operator: $\langle\Psi(t_{\rm f}) \vert \hat \eta\vert \Psi(t_{\rm in})\rangle=\langle\Psi \vert {\rm e}^{-i\hat H\Delta t}\hat \eta \vert\Psi\rangle= {\rm e}^{-iE_{\Psi}\Delta t}$, where $E_{\Psi}$ is the expectation value of the Hamiltonian operator $\hat H$ for the state $\Psi$ (in the Heisenberg picture) and $\Delta t=t_{\rm f}-t_{\rm in}$. It is well known that transition amplitudes exhibit a crossing symmetry; that is exchanging the initial and final states ($t_{\rm in}\leftrightarrow t_{\rm f}$) and taking $E_{\Psi}\to -E_{\Psi}$ leaves the transition amplitudes unchanged. We can use this crossing symmetry to reinterpret negative energy states. In particular, a negative energy particle in an initial (final) state can be interpreted as a positive energy antiparticle in a final (initial) state \cite{Dhar:1969gb}. This way we only deal with positive energy initial and final states. For example, the vacuum decay process discussed above becomes reinterpreted as the vacuum polarization process. We would like to stress, however, that crossing symmetry is a symmetry of transition amplitudes and not the quantum theory itself. Therefore it can not be used to completely remove negative energy states within the time-symmetric quantization. The situation here is different from the standard quantization scheme of systems with gauge invariance. The gauge symmetry is a fundamental symmetry of a theory and thus negative energy or norm states are redundant, unphysical states that can be completely removed by fixing the gauge.   

Despite the somewhat ad hoc and non-dynamical nature of the reinterpretation postulate, it is clear that it does the job ``phenomenologically" by removing apparent pathologies associated with the negative energy states in a generic interacting theory. This follows from the fact that the Hamiltonian in the above-discussion is not restricted to be a free-Hamiltonian. In fact, what one needs for the consistency of the theory is to avoid asymptotic \emph{in} and \emph{out} negative states and that is exactly established by the reinterpretation postulate. Therefore, together with the automatic causality condition (\ref{15}), the reinterpretation postulate implies that the time-symmetric quantization formalism may be considered as a valid alternative to the standard quantization scheme. 
   
\section{Discussion and conclusion }

It is widely believed that Hawking radiation provides a consistent thermodynamic interpretation of Bekenstein's black hole entropy \cite{Bekenstein:1972tm}. From this point of view the absence of such thermal radiation in the time-symmetric quantization may seem like a major deficiency. We would like to point out that, while the relation between the Hawking radiation and black hole entropy is suggestive, there is no rigorous physical proof of such a relation. Moreover, the physical origin of the two phenomena are completely different: Bekenstein entropy is based on entirely classical solutions to the Einstein equations, while the thermal radiation can be seen as a purely kinematic quantum mechanical effect, as discussed in Section 1. In truth, the thermodynamic interpretation of gravity may ultimately require understanding at a deeper level; the result of our paper is simply intended to elucidate the origins of the associated black hole information paradox.

In conclusion, we have argued that thermal radiation does not appear when the fields are quantized in spacetimes with horizons within the time-symmetric quantization formalism. The thermal effects are simply cancelled out between the positive and negative energy contributions. We have also shown that the time-symmetric quantization is causal and deals only with positive norm states. Finally we have argued for the reinterpretation postulate in order to have a sensible quantum theory with negative energy states. Accordingly the time-symmetric formalism in the free field limit is a consistent theory in which the black hole information paradox does not arise, implying that a time-asymmetric quantization is indeed a necessary condition for its existence.

\paragraph{Acknowledgements.} We are grateful to Ray Volkas for reading the manuscript and making comments. The work was partially supported by the ARC. 

%\newpage
%\baselineskip=8pt

\end{document}